\documentclass[fleqn,usenatbib]{mnras}
\usepackage{CJK,float}
\usepackage{newtxtext,newtxmath}

\usepackage[T1]{fontenc}

\DeclareRobustCommand{\VAN}[3]{#2}
\let\VANthebibliography\thebibliography
\def\thebibliography{\DeclareRobustCommand{\VAN}[3]{##3}\VANthebibliography}

\usepackage[dvipsnames]{xcolor}

\usepackage{graphicx}	
\usepackage{amsmath}	
\usepackage{hyperref}
\usepackage{fontawesome}

\makeatletter
\newcommand{\github}[1]{%
   \href{#1}{\faGithub}%
}
\makeatother

\title[Galaxy Deconvolution with Unrolled Plug-and-Play ADMM]{Galaxy Image Deconvolution for Weak Gravitational Lensing with Unrolled Plug-and-Play ADMM}

\author[T. Li et al.]{
Tianao Li ({李天骜}),$^{1}$\thanks{E-mail: lukeli0425@gmail.com}
Emma Alexander$^{2}$\thanks{E-mail: ealexander@northwestern.edu}
\\
$^{1}$Department of Electronic Engineering, Tsinghua University, Beijing 100084, P.R.China \\
$^{2}$Department of Computer Science, Northwestern University, Evanston, IL 60208, USA
}

\date{Accepted XXX. Received YYY; in original form ZZZ}

\pubyear{2022}

\begin{document}
\begin{CJK}{UTF8}{gbsn}
\label{firstpage}
\pagerange{\pageref{firstpage}--\pageref{lastpage}}
\maketitle
\end{CJK}
\begin{abstract}
Removing optical and atmospheric blur from galaxy images significantly improves galaxy shape measurements for weak gravitational lensing and galaxy evolution studies. This ill-posed linear inverse problem is usually solved with deconvolution algorithms enhanced by regularisation priors or deep learning. We introduce a so-called "physics-informed deep learning" approach to the Point Spread Function (PSF) deconvolution problem in galaxy surveys. We apply algorithm unrolling and the Plug-and-Play technique to the Alternating Direction Method of Multipliers (ADMM), in which a neural network learns appropriate hyperparameters and denoising priors from simulated galaxy images.
We characterise the time-performance trade-off of several methods for galaxies of differing brightness levels {\color{black}as well as our method's robustness to systematic PSF errors and network ablations. We show an improvement in reduced shear ellipticity error of 38.6\% (SNR=20)/45.0\% (SNR=200) compared to classic methods and 7.4\% (SNR=20)/33.2\% (SNR=200)} compared to modern methods.
\github{https://github.com/Lukeli0425/Galaxy-Deconv}

\end{abstract}




\section{Introduction}
\label{sec:1}
Images captured by real-world systems are degraded by nonidealities in the atmosphere, optics, and sensors. Processing images to correct these effects can significantly reduce errors in physical measurements drawn from astronomical images, such as the galaxy shape estimation used in cosmological weak lensing analysis to advance our understanding of cosmological models, the relationship between visible and dark matter, and cosmological-scale gravity~\citep{kaiser1994method,bartelmann2001weak,mandelbaum2015great3, mandelbaum2018weak}. Conventionally, these nonidealities are modelled by convolution with a blur kernel called the Point Spread Function (PSF), followed by noise. Many computer vision algorithms have been developed to solve this ill-posed linear inverse problem, known as deconvolution. Here, we characterise the improvements that can be made to shear estimation through several deconvolution approaches, with a focus on a data-driven unrolled optimisation.

Two classic deconvolution algorithms still in widespread use are inverse filtering, such as Fourier division, and the iterative \textit{Richardson-Lucy algorithm}~\citep{richardson1972bayesian, lucy1974iterative}, which optimizes the image estimate using gradient descent with an adaptive step size. {\color{black}These methods can amplify image noise, highlighting the importance of parametric models for stable estimation of galaxy shape, such as the Fourier Power Function Shapelets method~\citep{li2022analytical}, which uses a pipeline of five shape estimation steps based on shapelet representation and bias correction.}

With the recent flourishing of deep learning, many new galaxy image deconvolution algorithms have been developed. In computer vision, so-called physics-informed machine learning combines an explicit forward model of the image formation process with data-driven regularisation. Notably, \cite{sureau2020deep} introduced two such deconvolution approaches enhanced by deep learning: \textit{Tikhonet}, which post-processes the results of Tikhonov deconvolution with a deep neural net, and \textit{ADMMNet}, an iterative deconvolution framework with learned denoising parameters. {\color{black} \cite{nammourshapenet} then proposed \textit{ShapeNet}, which retrains the Tikhonet network with a loss function designed to preserve galaxy shape.} ADMMNet is based on the Alternating Direction Method of Multipliers \citep[ADMM,][]{boyd2011distributed}, which finds the maximum-a-posteriori estimate of the ground truth image by alternating between tractable subproblems. After each subproblem, estimates are updated with a variable step size, where the optimal step size depends on the generally-unknown prior distribution of the dataset. ADMMNet takes advantage of the Plug-and-Play approach~\citep{chan2016plug}, in which the subproblems are seen to correspond to alternating deconvolution and denoising steps. This modular structure is exploited by including an off-the-shelf denoiser (typically a deep neural network) in place of the standard denoising step, for more accurate data-driven priors within the context of an interpretable and well-studied optimisation problem. {\color{black}In the case of ADMMNet, the deconvolution step is untrained and the denoising network is trained to remove additive noise from unblurred images, then the alternating steps are applied for a freely-varying number of iterations until convergence.}

\begin{figure*}
    \includegraphics[width=2\columnwidth]{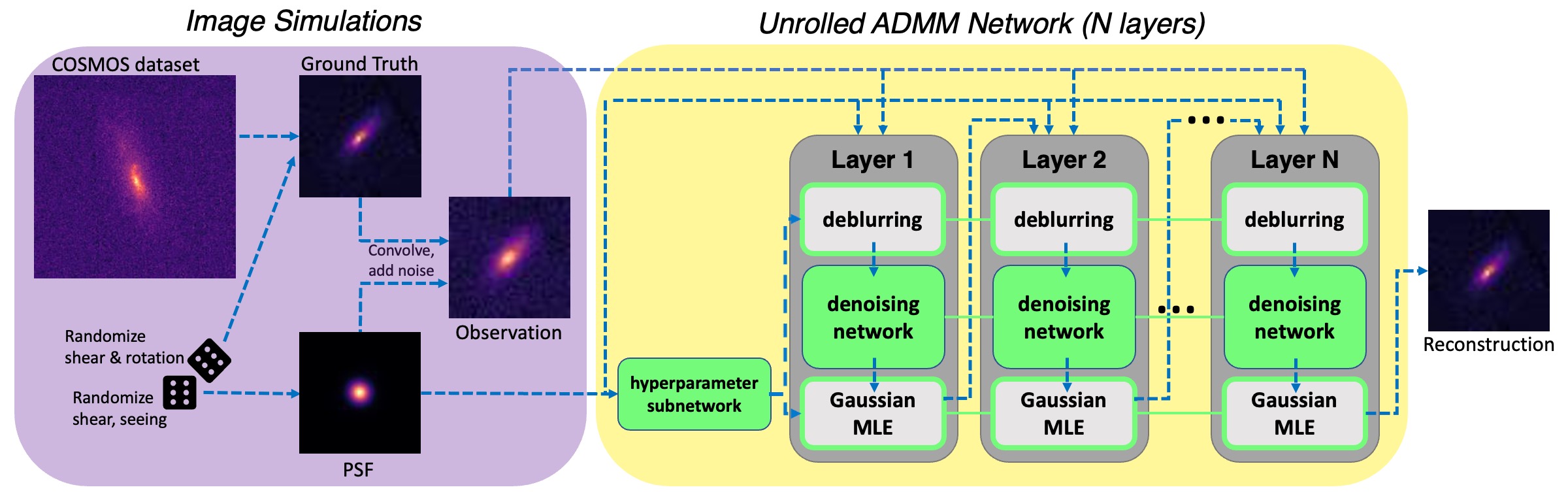}
    \caption{\textbf{Overview of the image processing pipeline.} \emph{Image simulations}: we simulated galaxy images for the upcoming LSST ground-based sky survey by modifying galaxy images from the COSMOS Dataset with the modular galaxy image simulation toolkit Galsim. The raw images were randomly sheared and rotated to simulate weak lensing effects. These images were convolved with randomised atmospheric and optical PSFs and scaled to varying brightnesses before Gaussian noise was added. Finally, all images were downsampled to the pixel scale of LSST. \emph{Model}: we unrolled Plug-and-Play ADMM with a Gaussian likelihood into an N-layer neural network. The proposed method takes N=8 but separate networks were trained for N = 2 and 4 for full characterization of the method. Light grey boxes denote fixed computations while green boxes are neural networks (a small convolutional neural network for hyperparameter estimation and a ResUNet for denoising). Green outlines indicate the use of predicted hyperparameters in the fixed computations and horizontal green lines show that weights are shared across network layers. The network was trained with multiscale L1 loss but reconstruction quality is analysed using reduced shear error.}
    \label{fig:pipeline}
\end{figure*}

{\color{black}We propose an \textit{Unrolled} Plug-and-Play ADMM network}. In an "unrolled" or "unfolded" iterative optimisation, the iteration count is fixed a priori and each layer of a neural network performs one stage of the optimisation~\citep{monga2021algorithm}. Layers contain fixed computations reflecting built-in knowledge (the image formation model) as well as optimisation parameters (the deep denoising network and a small number of step size hyperparameters) that are treated as learnable weights. This approach allows efficient, end-to-end training of a limited number of weights and a predictable compute time.  
Fig.~\ref{fig:pipeline} shows the proposed network architecture, with fixed computations shown in light grey and learned parameters indicated in green. The green boxes are neural networks, green outlines denote the use of trained hyperparameters, and horizontal green lines indicate that learned weights are identical across layers. {\color{black}Details of network structure follow \cite{sanghvi2022photon}, including a subnetwork for parameter estimation so that the network can respond to the image noise level and PSF, though we use a Gaussian-noise-based update step in place of the Poisson step. We propose an 8-iteration network but report performance for 2- and 4-iteration networks as well.}

In this paper, we benchmark classical and modern deconvolution methods on realistic simulated data for ellipticity error, compute time, and sensitivity to PSF errors, showing that unrolled Plug-and-Play ADMM, not previously applied to astronomical images, outperforms former methods robustly. We also provide source code and trained weights of the suggested model. Our user-friendly framework allows one to easily simulate their own data set and train the model under their settings. Our code and tutorials are all available on our GitHub repository.

\begin{figure*}
	\includegraphics[width=2\columnwidth]{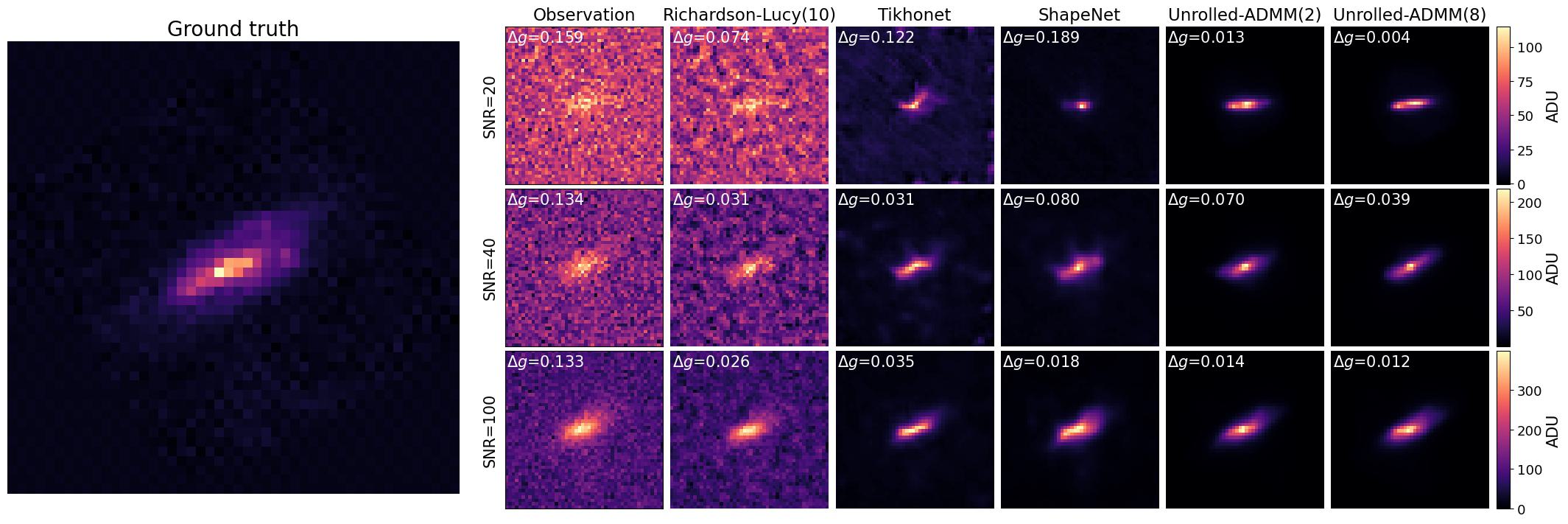}
    \caption{\color{black}\textbf{An illustration of performance at multiple SNR levels.} The size of each image is 48x48 pixels and each row corresponds to a different SNR level (see brightness in ADU on color bars to the right). Ellipticity errors are reported for each image in terms of reduced shear, using robust shear estimator FPFS. Columns from left to right: simulated observation, Richardson-Lucy deconvolution with 10 iterations, Tikhonet, ShapeNet, and unrolled ADMM with 2 and 8 iterations. Methods gradually fail with the decrease of SNR but the proposed unrolled ADMM remains effective across SNR levels. For typical performance of each method across galaxy samples, see later figures.}
    \label{fig:grid}
\end{figure*}

\vspace{-0.6cm}
\section{Experiments}
\subsection{Image Simulations}
To benchmark classical methods and train our network, clean ground truth images must be paired with realistic PSFs and noisy blurred images. We created our dataset by simulating ground-based observations with the modular galaxy image simulation toolkit Galsim\footnote{\href{https://github.com/GalSim-developers/GalSim}{https://github.com/GalSim-developers/GalSim}}
~\citep{ROWE2015121} and the COSMOS Real Galaxy Dataset~\citep{mandelbaum_rachel_2012_3242143}. 

The image simulation process is illustrated in Fig.~\ref{fig:pipeline} and was based on the upcoming Legacy Survey of Space and Time~\citep[LSST,][]{ivezic2019lsst}. Raw images were picked randomly from the COSMOS dataset and then randomly sheared (uniform distribution of $[0.01,0.05]$) and rotated (uniform distribution of $[0,2\pi]$) to simulate weak lensing effects. We then applied both atmospheric and optical blurring. The atmospheric PSFs were simulated with the Kolmogorov function, whose seeing (FWHM) was sampled from {\color{black}a distribution~\citep{connolly2010simulating} as used in the GREAT3 Challenge~\citep{mandelbaum2014third}.} To simulate the atmospheric variation that leads to shape noise, we added shear (uniform distribution of $[0.01,0.03]$) and rotation (uniform distribution of $[0,2\pi]$) to the Kolmogorov functions. Our optical PSFs were simulated using Galsim, using the LSST diameter for diffraction and built-in ranges for obscuration, defocus, coma aberration and astigmatism. The two PSF components were convolved to get the overall PSF, which is saved for training and testing. The clean galaxy image (ground truth) was then convolved with this simulated PSF, {\color{black}then the brightness was scaled to fit the SNR of interest (SNR=$\sqrt{\Sigma I(x)^2/\sigma^2}$). Specifically, based on LSST read noise and a sky level of 350 ADUs/pixel\footnote{\href{https://smtn-002.lsst.io}{https://smtn-002.lsst.io}}, we added Gaussian noise with a standard deviation $\sigma=19.4$ ADUs.} Finally, all images were downsampled to meet the pixel scale of the LSST (0.2 arcsec).
{\color{black} While these images are mean-subtracted, a significant difference between our dataset and the GREAT3 simulations is that we do not normalise image brightness, so the hyperparameter subnetwork can readily adapt our method in response to SNR based on input brightness.}

\vspace{-0.4cm}
\subsection{Network Training}
We train on 40,000 samples using multi-scale L1 loss between ground truth and reconstructed images for 50 epochs with the Adam optimizer on an NVIDIA GeForce RTX 2080 Ti GPU. A separate network was trained for each iteration count (N=2,4,8). ResUNet~\citep{He_2016_CVPR} serves as the plugin denoiser (17,007,744 parameters), introducing skip connections in the neural network to avoid the problem of vanishing or exploding gradients as the network gets deeper with the increase of iterations. A CNN (80,236 parameters) was trained to learn the step size hyperparameters. We use the open source machine learning framework PyTorch
, with which we also provide tools for dataset generation and benchmarking.

\vspace{-0.3cm}
\subsection{Performance across SNR levels}
\label{sec:err-snr}
\begin{figure}
	\includegraphics[width=\columnwidth]{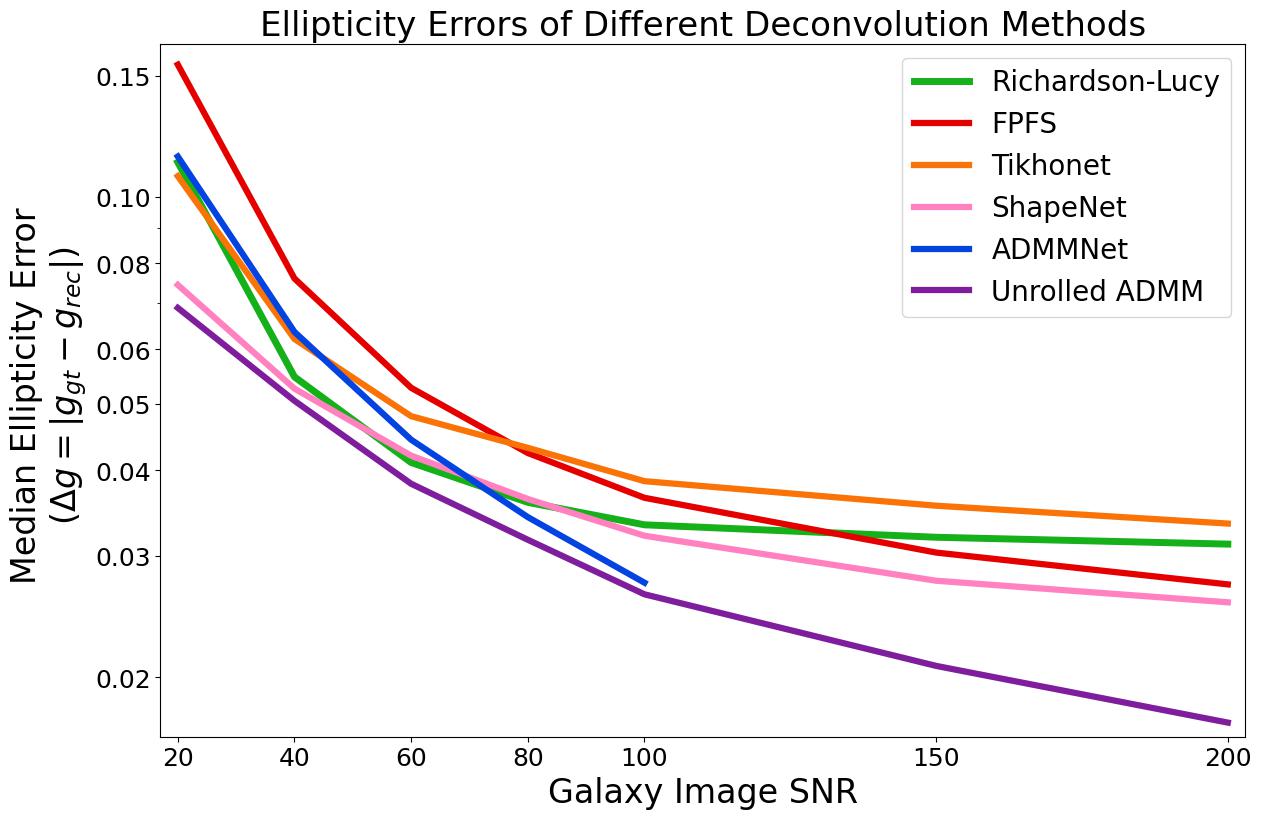}
    \caption{\color{black}\textbf{Median ellipticity errors at multiple SNR levels.} As all the methods gradually fail with the decrease of SNR, our 8-iteration unrolled ADMM network presents the best performance across SNR levels. ADMMNet errors are reported from {\protect\cite{sureau2020deep}}, Tikhonet and ShapeNet were reimplemented and retrained on our dataset. All deconvolution methods were followed by FPFS-based parametric shape estimation with a single-pixel PSF, except the reported FPFS errors which used the true PSF for inverse filtering. For standard errors of each method, please see Fig.~\ref{fig:error_time}.}
    \label{fig:err-snr}
\end{figure}

\begin{figure}
	\centering
	\includegraphics[width=\columnwidth]{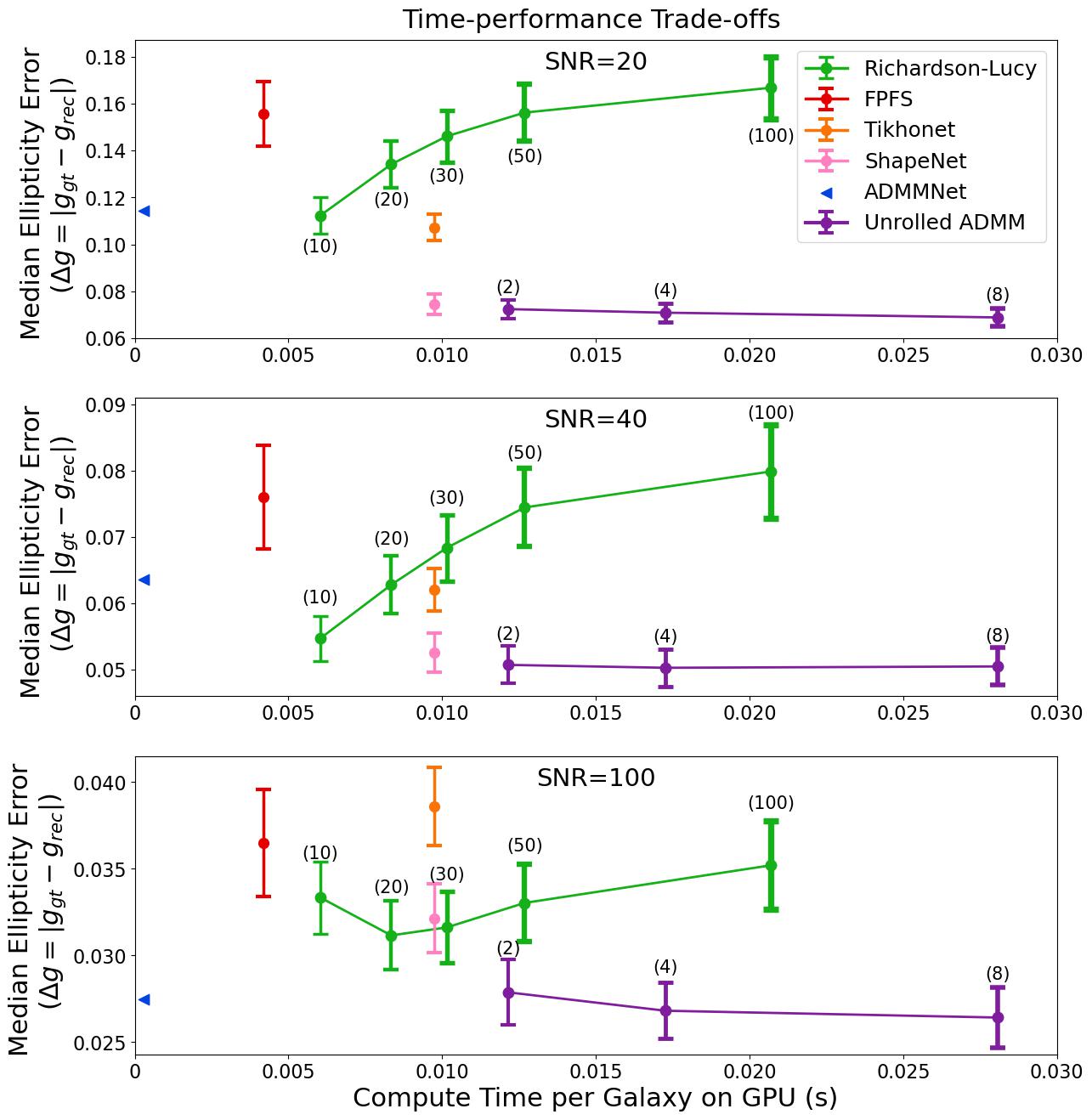}
    \caption{\color{black}\textbf{Time-performance trade-offs at representative SNRs.} The proposed method (unrolled ADMM with 8 iterations) shows higher performance and lower variance than other methods across SNR levels, at the cost of more compute time. ADMMNet errors are shown as reported by the authors with blue markers on the y-axis; reported compute time per galaxy was 8.75s for this method. The three panels correspond to galaxy image SNR of 20, 40 and 100. Error bars show 5x standard error and numbers in parentheses indicate iteration counts.
    }
    \label{fig:error_time}
\end{figure}
We tested the suggested method at different SNR levels and compared its performance to other algorithms. Sample images deconvolved by each of the algorithms are shown in Fig \ref{fig:grid}. At low SNR, the Richardson-Lucy algorithm results in noisier reconstructions in which the galaxy is gradually submerged in the noisy patterns in the background. {\color{black}Modern methods suppress background noise but can distort galaxy shape, particularly at low SNR. We use the FPFS toolkit for ellipticity estimation, in conjunction with the known PSF for inverse filtering (labelled as the \textit{FPFS} method) or with a single-point PSF as a stable estimator without inverse filtering, which we apply to the output of all reported deconvolution methods throughout the paper. Errors are reported in terms of median reduced shear~\citep[$\Delta g$,][]{bartelmann2001weak}.} 

\begin{figure*}
	\centering
	\includegraphics[width=2\columnwidth]{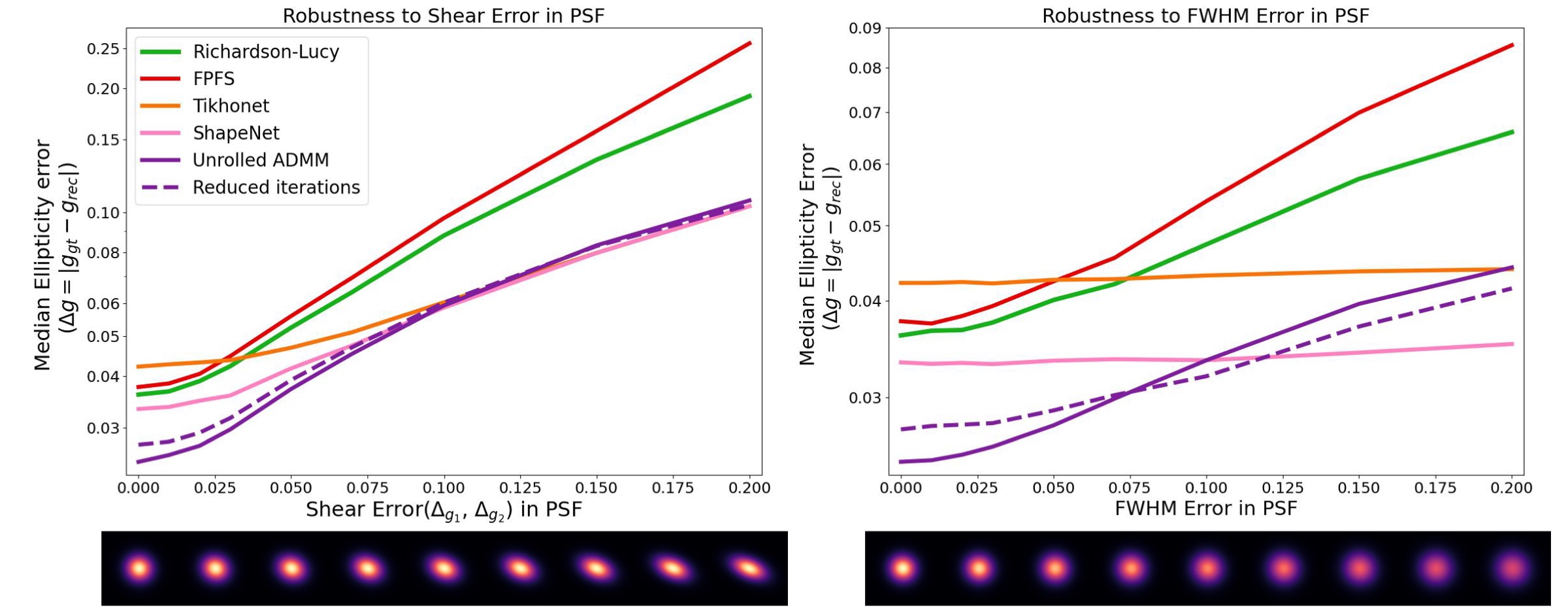}
    \caption{\color{black}\textbf{Ellipticity error due to systematic errors in the assumed PSF.} The left panel corresponds to shear error in the assumed PSF while the right corresponds to size (FWHM, i.e. the atmospheric seeing) error in the PSF. The cartoon bars under the horizontal axes give a visual indication of the systematic error in PSF. The proposed method and classical methods show higher sensitivity to model mismatch. Note that under high PSF error, the proposed method can be matched and even outperformed by the modern methods tested. This can be mitigated slightly by reducing the iterations, with a two-iteration version of the proposed method shown in dashed purple.\vspace{-.2cm}}
    \label{fig:psf_err}
\end{figure*}
We characterise the reduced shear ellipticity error of each method across a dataset of 10,000 simulated galaxy images for 7 SNR levels from 20 to 200, shown in Fig.~\ref{fig:err-snr}. {\color{black}We tested the Richardson-Lucy algorithm with 10 iterations, the built-in robust Fourier division method of FPFS, and a reimplementation and retraining of Tikhonet and ShapeNet on our dataset, and we also include reported errors from ADMMNet for comparison.} The proposed method consistently outperforms classical and modern methods across SNR levels.

\vspace{-0.3cm}
\subsection{Time-Performance Trade-off}
{\color{black} The improved performance of the proposed method comes at the cost of elevated computational demand. Fig.~\ref{fig:error_time} shows the trade-off between compute time and ellipticity error at SNR levels of 20, 40 and 100 on the test dataset. Error bars show 5x the standard error of the ellipticity errors. Reducing the iterations in the proposed method reduces the compute time per galaxy at the cost of slightly increased error (purple). This is not a universal trade-off for iterative methods; increasing the number of iterations of the Richardson-Lucy algorithm boosts noise and degrades performance on dim galaxies (green). Performance for ADMMNet (blue) is as originally reported by the authors, who also report an average time per galaxy of 8.75 seconds, orders of magnitude higher than the other methods considered here. This is not likely due to hardware differences, as our GPU roughly matches their reported timing for Tikhonet. Rather, ADMMNet is allowed to iterate until convergence rather than undergoing a prefixed number of steps, enabling dramatically extended computation.}

\vspace{-0.25cm}
\subsection{Robustness to Systematic Errors in PSF}
In real-world observations, PSFs at the center of the galaxies are calculated through interpolation from PSFs measured at stars, yet the spatial variance of the PSF across the field of view is often difficult to model~\citep{https://doi.org/10.48550/arxiv.2209.05489}. As a result, interpolation cannot precisely reconstruct the effective PSF across galaxies. This requires the subsequent deconvolution algorithm to be robust to systematic errors in the PSFs. 

Errors in PSF shape and size result in additive and multiplicative shear biases~\citep{mandelbaum2018weak}. In our PSF model, these two types of systematic errors correspond to errors in the PSF shear and the full-width half maximum (FWHM) of the Kolmogorov function, i.e., the atmospheric seeing. To test the robustness of the tested methods to PSF noise, we characterized the ellipticity errors caused by systematic errors (i.e. model mismatch) between the PSF used for image formation and the PSF provided to the deconvolution algorithms. Fig.~\ref{fig:psf_err} shows these errors, with an illustration of the type and level of error shown in the cartoon PSFs below the plots. Under reasonable PSF errors, the proposed method maintains best performance, but other modern methods are more robust to large mistakes in PSF shape and size. A 2-iteration version of the proposed method (dashed purple) is less sensitive to large model mismatch, showing that the iteration number for the proposed method should be reduced when high errors in PSF estimation are expected. 

\vspace{-0.4cm}
{\color{black}
\subsection{Ablation Studies}
\begin{figure}\centering
    \includegraphics[width=\columnwidth]{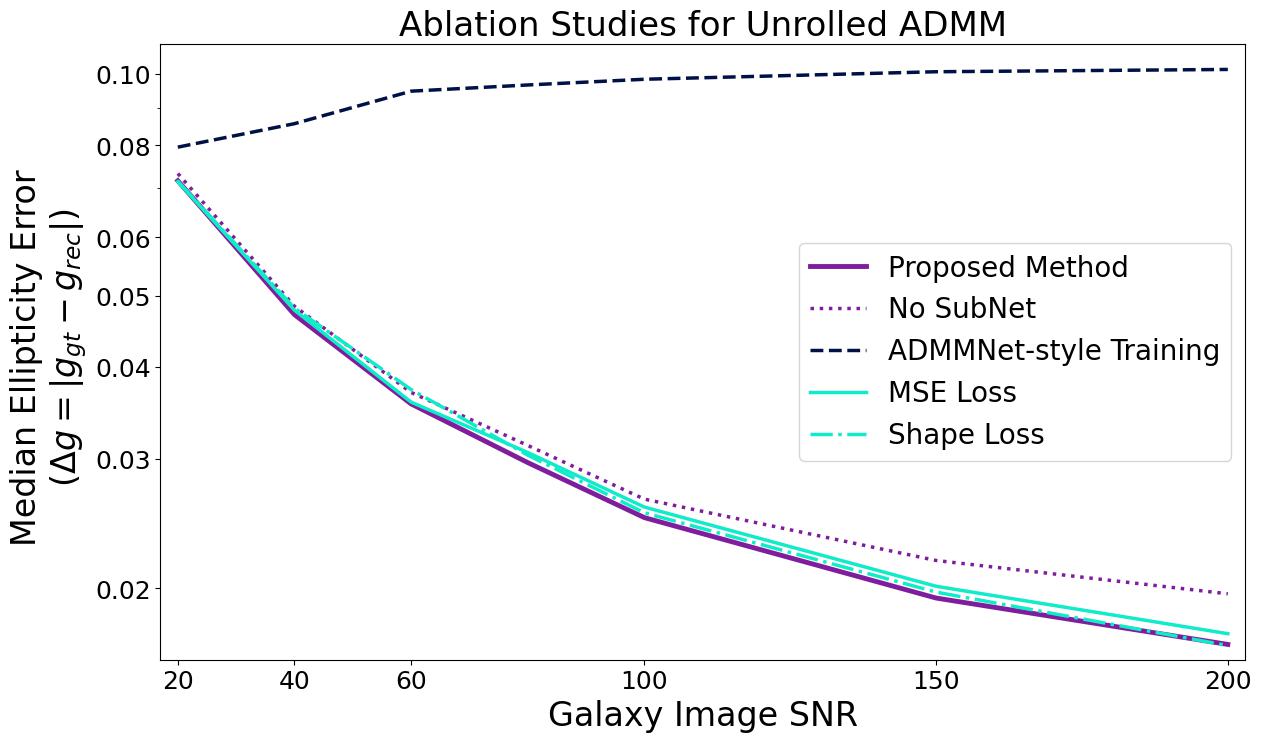}
    \caption{\color{black}\textbf{Ablations isolate contributions of network and training design decisions.} Changing the loss function from the proposed multiscale loss to shape loss (as in Shapenet, dashed teal) or MSE (as in Tikhonet, solid teal) only slightly reduces performance. Removing the hyperparameter subnetwork (dashed purple) or later iteration layers (shown in Fig.~\ref{fig:error_time}) reduces performance at high SNR. Training the denoiser but not the deconvolution (ADMMNet-style training) reduces 
    performance dramatically.\vspace{-0.5cm}}
    \label{fig:ablation}
\end{figure}

By altering the structure of the network and retraining, we can isolate the effects of individual design choices to the proposed method. Fig.~\ref{fig:ablation} shows the effects of 3 types of alterations. Changes in the loss function are shown in teal. Changing from multiscale loss to MSE loss, as used in Tikhonet, reduces performance slightly, and to shape loss, as used in Shapenet, leads to  comparable performance. Removal of the subnetwork that predicts hyperparameters (dashed purple), as in Tikhonet, Shapenet, and ADMMNet, reduces performance at high SNR, as does removal of iteration layers (see Fig.~\ref{fig:error_time}). ADMMNet-style training (denoiser-only training on noised images without PSFs, shown in dark blue) leads to a considerable loss in performance. It should be noted that ADMMNet is allowed to iterate to convergence, so that our fixed-iteration network structure does not reflect identical computation.}

\vspace{-0.55cm}
{\color{black}
\section{Discussion}
The proposed method could be further improved in several ways. While we have primarily tested deconvolution algorithms in a non-blind (known-PSF) setting, in practice the PSFs cannot be known exactly and the process is at least partially blind. We present an initial characterisation of performance in response to simple errors in PSF, but higher order errors must be considered and addressed for best performance. Our simulated data can be improved with more accurate optical models for the LSST PSF~\citep{https://doi.org/10.48550/arxiv.2209.05489} and more realistic generative models such as ~\cite{peterson2005lsst}, which will improve further once observations begin. Additionally, to address spatial variations, PSF interpolation can be included in the pipeline and combined with low-rank deconvolution used in \cite{farrens2017space} and \cite{yanny2020miniscope3d}. This may be of particular interest for larger galaxies with fine internal structure that could be resolved using high-quality deconvolution. It would also be interesting to test the effect of training set size on the generalizability of the methods considered here. The shared weights across iteration layers in a unrolled network may confer an advantage in a low-training-data setting, in which parameter-efficient methods would be able to take advantage of data from novel instruments earlier in their collection periods.}

\vspace{-0.55cm}
\section{Conclusions}
We have simulated galaxy images to train a specific subset of the parameters in a well-studied optimisation method. In comparison with traditional and learned methods, the suggested model is more robust and effective, because it combines a powerful and well-analysed iterative method with data-driven priors learned by an end-to-end optimised deep neural network through algorithm unrolling. Previous work on data-driven ADMM, ADMMNet, trained only the denoising step and required a varying number of iterations to reach convergence, leading to orders of magnitude more time per galaxy to compute. Our fixed iteration count and joint training across both subproblems achieves better performance in much less time.

We tested the suggested method on data simulated with realistic galaxy images, and compared its performance to classical and modern methods in terms of the median and variance of reduced shear ellipticity error over a wide range of galaxy brightnesses, as well as the time consumption, the robustness to systematic PSF noise, and the influence of network ablations and iteration count. Our experiments showed that 8-iteration unrolled Plug-and-Play ADMM provides the most accurate ellipticity measurements across SNR levels at the cost of slightly elevated compute time. 

Because image statistics and computational resources vary across applications and observational conditions, the networks we trained will not be the best method for every galaxy image deconvolution problem. Our simulations were modelled on a specific instrument currently in development, and deviations from these parameters will reduce performance. Because of this, we provide an open-source, user-friendly code base for generating and retraining on simulated datasets for astronomical deconvolution problems. Simulation results allow researchers to develop and select the most suitable algorithm for their performance and speed needs.

\vspace{-0.5cm}
\section*{Acknowledgements}
We would like to thank Adam Miller, Jason Wang, Keming Zhang, Nick Antipa, and an anonymous reviewer for their valuable advice, and the Computational Photography Lab (esp. Aniket Dashpute) at Northwestern University for support of computational resources.

\vspace{-0.5cm}
\section*{Data Availability}

Our GitHub repository (\href{https://github.com/Lukeli0425/Galaxy-Deconv}{https://github.com/Lukeli0425/Galaxy-Deconvolution}) includes the code needed to regenerate the simulated data, as well as a link to our dataset. We also provide the suggested unrolled ADMM model with trained weights under LSST settings along with tutorials for using it, allowing a convenient deconvolution with the suggested method for further research.


\vspace{-0.4cm}
\bibliographystyle{mnras}
\bibliography{citations}

\bsp	
\label{lastpage}
\end{document}